\documentclass[prl,twocolumn,superscriptaddress,showpacs]{revtex4}
\usepackage{amsmath}
\usepackage{bm}
\usepackage{graphicx}
\usepackage{amssymb}
\usepackage{color}
\usepackage{times}
\begin{document}
%%%%%%%%%%%%%%%%%%%%%%%%%%%%%%%%%%%%%%%%%%%%%%%%%%%%%%%%%%%%%%%%%%%%%%%%%%
% comment out for single column
%\twocolumn[%
%\hsize\textwidth\columnwidth\hsize\csname@twocolumnfalse\endcsname
%\draft
%%%%%%%%%%%%%%%%%%%%%%%%%%%%%%%%%%%%%%%%%%%%%%%%%%%%%%%%%%%%%%%%%%%%%%%%%%
\newcommand{\figwidth}{0.93\columnwidth}
\newcommand{\ffigwidth}{0.4\columnwidth}
%%%affiliations
\newcommand{\warwick}{Department of Physics and Centre for Scientific Computing, University of Warwick, Coventry,
  CV4 7AL, United Kingdom}
\newcommand{\usal}{Departamento de Fisica Fundamental, Universidad de Salamanca, 37008 Salamanca, Spain}
\title{Multifractal analysis with the probability density function\\ at the three-dimensional Anderson transition}
\author{Alberto Rodriguez}
\email[Corresponding author: ]{A.Rodriguez-Gonzalez@warwick.ac.uk}
\affiliation{\warwick}
\affiliation{\usal}
\author{Louella J. Vasquez}
\affiliation{\warwick}
\author{Rudolf A. R\"omer}
\affiliation{\warwick}
\date{$Revision: 1.68 $, compiled \today}
%\date{\today}
%
\begin{abstract}
The probability density function (PDF) for critical wavefunction amplitudes is studied in the three-dimensional Anderson model.  We present a formal expression between the PDF and the multifractal spectrum $f(\alpha)$ in which the role of finite-size corrections is properly analyzed.  We show the non-gaussian nature and the existence of a symmetry relation in the PDF.
From the PDF, we extract information about $f(\alpha)$ at criticality such as the presence of negative fractal dimensions and we comment on the possible existence of termination points. A PDF-based multifractal analysis is hence shown to be a valid alternative to the standard approach based on the scaling of general inverse participation ratios.
\end{abstract}
\maketitle
%%%%%%%%%%%%%%%%%%%%%%%%%%%%%%%%%%%%%%%%%%%%%%%%%%%%%%%%%%%%%%%%%%%%%%%%%%
%% main text
%%%%%%%%%%%%%%%%%%%%%%%%%%%%%%%%%%%%%%%%%%%%%%%%%%%%%%%%%%%%%%%%%%%%%%%%%%
The fluctuations and correlations of wave amplitudes are of primary importance for the understanding of many classical and quantum systems. This is arguably most pronounced in the physics of Anderson localization \cite{EveM08}. Here, recent advancements in theory \cite{MirFME06,ObuSFGL08,EveMM08a}, experiments in classical \cite{HuSPS08} and quantum waves \cite{BilJZB08,RoaDFF08,HasSMI08} as well as numerical methods \cite{RodVR08,VasRR08a} have led to unprecedented insights into the nature of the localization-delocalization transition.
In contrast to the weak- or strong-disorder limits where the description of nearly-extended or strongly localized states is well-known, e.g.\ from random matrix theory \cite{MulMMS97}, the intensity distribution at the metal-insulator transition is more involved due to the multifractal nature of the states \cite{Aok83,EveM08,LudTED05}. 
The possibility of carrying out a multifractal analysis directly from the raw statistics of intensities $\vert\psi\vert^2$, i.e.\ the probability density function (PDF), is especially interesting since their distributions can be measured experimentally in classical \cite{HuSPS08} and quantum \cite{HasSMI08,MorKMW03,MorKMG02} experiments. The PDF at criticality is closely related to the multifractal spectrum $f(\alpha)$. However, the numerical relation between the PDF and $f(\alpha)$ has not been completely elucidated. In this Letter we show how to obtain the multifractal spectrum based on the PDF. The PDF-to-$f(\alpha)$ connection is a numerically much simpler procedure than the usual scaling of $q$-moments of $\vert\psi\vert^2$ \cite{Jan94a}. Furthermore, it yields direct understanding of physical properties at criticality, such as the existence of a symmetry relation, the observation of negative fractal dimensions and the physical meaning of the possible appearance of termination points. We apply the PDF-based approach to the three-dimensional Anderson model within the Gaussian orthogonal ensemble, using a large number of critical states at $E=0$ and very large system-sizes up to $L^3=240^3$ \cite{VasRR08a}.

At criticality, the intensity distribution has the scaling form
\begin{equation}
 \mathcal{P}_L(\vert\psi\vert^2)\sim\left(1/\vert\psi\vert^2\right)L^{f\left(-\ln \vert\psi\vert^2/\ln L\right)-d}.
 \label{eq:pdfpsi}
\end{equation}
In terms of the variable $\alpha\equiv -\ln \vert\psi\vert^2/\ln L$, the PDF is $\mathcal{P}_L(\alpha)\sim L^{f(\alpha)-d}$, where $f(\alpha)$ is the multifractal spectrum, i.e.\ the fractal dimensions of the different $\alpha$-sets made up of the points where $|\psi_i|^2=L^{-\alpha}$. The standard $f(\alpha)$ spectrum is usually constructed by a Legendre transformation \cite{Jan94a} of the scaling exponents $\tau(q)$ for the  generalized inverse participation ratios $L^d\left<\vert\psi_i\vert^{2q}\right> \sim L^{-\tau(q)}$ \cite{MilRS97,MirFME06,ObuSFGL08,EveMM08a,VasRR08a,RodVR08}.

\begin{figure}
  \centering
   \includegraphics[width=\figwidth]{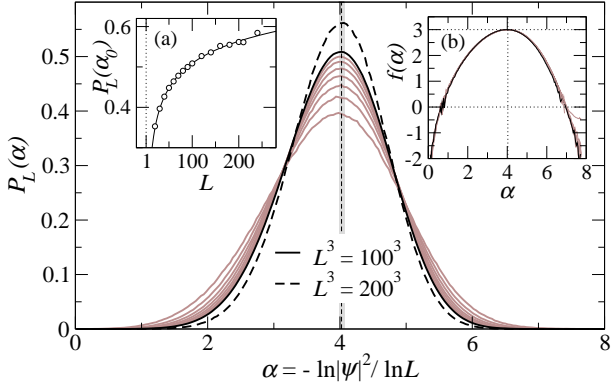}
   \caption{(color online) PDF at criticality for $\Delta\alpha=0.04$. The gray lines correspond to $L$ from $30$ (bottom) to $90$ (top). Standard deviations are within the line width. For $L\leqslant 100$ and $L>100$ we average over $2.5\times 10^4$ and $100$ states respectively. The vertical dashed line marks the mean value for $\alpha_0=4.024\pm0.022$ using $L$ from $50$ to $200$. The shaded vertical region corresponds to its $95\%$ c.i. Inset (a) shows the maximum values $\mathcal{P}_L (\alpha_0)$ vs $L$. Standard deviations are contained within symbol size. The solid line is the fit $a \ln(L/l)^b$, with $a=0.298\pm0.002$, $b=0.489\pm0.006$. Inset (b) shows the collapse of all the PDF from $L=30$ to $240$ onto the $f(\alpha)$.}
\label{fig-multipdf}
\end{figure}

The relation \eqref{eq:pdfpsi} between $\mathcal{P}_L(\alpha)$ and $f(\alpha)$ suggests a complete characterization of multifractality directly from the PDF. The proportionality in \eqref{eq:pdfpsi} contains an $L$-dependence which can be naively included as 
\begin{equation}
 \mathcal{P}_L(\alpha)=\mathcal{P}_L(\alpha_0) L^{f(\alpha)-d},
 \label{eq:pdfalpha}
\end{equation}
where $\alpha_0$ is the position of the maximum of the multifractal spectrum, $f(\alpha_0)=d$. Furthermore, since $f(\alpha) < d$ for all $\alpha\neq \alpha_0$, we see that $\alpha_0$ is in fact also the position of the maximum of the PDF itself. Hence $\mathcal{P}_L(\alpha_0)$ corresponds to the maximum value of the distribution, and $\alpha_0$ can be easily obtained numerically. 
As for $f(\alpha)$, the value of $\alpha_0$ obtained from the PDF must be $L$-invariant at criticality, as we show in Fig.~\ref{fig-multipdf}. The estimation for $\alpha_0=4.024\pm0.022$ from the PDF is in agreement with that obtained from gIPR scaling \cite{RodVR08}, $\alpha_0\in[4.024,4.030]$.
From the normalization condition we find $\mathcal{P}_L(\alpha_0)=\left(\int_0^\infty L^{f(\alpha)-d} d\alpha\right)^{-1}$. Using the saddle point method \cite{BleHan75}, justified in the limit of large $L$, we compute  $\mathcal{P}_L(\alpha_0)\sim \sqrt{\ln L}$, which holds very well even for small $L$ as shown in Fig.~\ref{fig-multipdf}(a). 
%%%%%%%%%%%%%%%%%%%%%%%%%%%%%%%%%%%%%%%%%%%%%%%%%%%%%%%%%%%%%%%%%%%%%%%%%%
%%%%%%%%%%%%%%%%%%%%%%%%%%%%%%%%%%%%%%%%%%%%%%%%%%%%%%%%%%%%%%%%%%%%%%%%%%
%
From the PDF for fixed $L$ the multifractal spectrum is hence straightforwardly obtained from \eqref{eq:pdfalpha} as 
 $f(\alpha)=d+\ln [\mathcal{P}_L(\alpha)/\mathcal{P}_L(\alpha_0)]/\ln L$. Alternatively, if $f(\alpha)$ is known, the PDF can be easily generated.   
We find excellent agreement between the singularity spectrum obtained from the PDF for $L=30,\ldots,240$ and the one obtained from the more involved box-size scaling of the gIPR \cite{VasRR08a}.
It must be emphasized that $\mathcal{P}_L(\alpha)$ is always system-size dependent, and it is through the $f(\alpha)$ spectrum that all PDFs for different $L$ collapse onto the same function, cp.\ Fig.~\ref{fig-multipdf}(b). Thus the $f(\alpha)$ can also be understood as the natural scale-invariant  distribution at criticality. 

\begin{figure}
  \centering
   \includegraphics[width=.9\columnwidth]{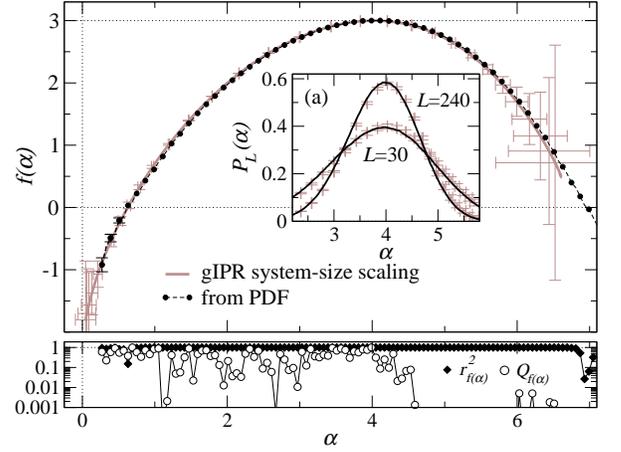}
   \caption{(color online) Comparison between $f(\alpha)$ obtained from system-size scaling of the PDF for $L\in[20,100]$ with $2.5\times10^4$ states for each $L$ and the spectrum obtained from the same set of data using system-size scaling of the gIPR. Only one every second symbol is shown for clarity. Standard deviations are within symbol size when not shown. The inset shows the numerical PDF (black) and the PDF calculated from the multifractal spectrum obtained from gIPR scaling (gray).  The bottom panel gives the values of the linear correlation coefficient $r^2$ and quality-of-fit parameter $Q$ for the $f(\alpha)$ obtained from the log-log linear fits of Eq.~\eqref{eq:correctedscaling}.}
\label{fig-giprsys}
\end{figure}

In order to minimize finite-size effects, we can also determine $f(\alpha)$ from the PDF using system-size scaling. We note that for a given $L$ the number of points per wavefunction with $\alpha\in[\alpha-\Delta\alpha/2,\alpha+\Delta\alpha/2]$ is $\mathcal{N}_L(\alpha)\equiv L^d \mathcal{P}_L(\alpha) \Delta\alpha$. Hence the following normalized volume of the $\alpha$-set $\widetilde{\mathcal{N}}_L(\alpha)\equiv L^d \mathcal{P}_L(\alpha)/\mathcal{P}_L(\alpha_0)$ obeying 
\begin{equation}
 \widetilde{\mathcal{N}}_L(\alpha)= L^{f(\alpha)},
 \label{eq:correctedscaling}
\end{equation}
can be used to extract $f(\alpha)$ from a series of systems with different $L$.
In Fig.~\ref{fig-giprsys} we compare the multifractal spectrum obtained using PDF scaling \eqref{eq:correctedscaling} with the one from gIPR scaling \cite{RodVR08} for $L\in[20,100]$ and having $2.5\times 10^4$ critical states for each size. We find very good agreement between both, as well as between the numerical PDFs and those generated from the gIPR $f(\alpha)$ [Fig.~\ref{fig-giprsys}(a)].

Numerically, the PDF is approximated by the histogram 
\begin{equation}
 \mathcal{P}_L(\alpha) \underset{\Delta\alpha\rightarrow 0}{\equiv} \left\langle \theta\left(\Delta\alpha/2-\left|\alpha+\ln|\psi_i|^2/\ln L\right|\right)\right\rangle/\Delta\alpha,
 \label{eq:histogram}
\end{equation}
where $\theta$ is the Heaviside step function and $\langle.\rangle$ involves an average over the volume of the system and all realizations of disorder. 
To minimize the uncertainty in the small $|\psi_i|^2$, which becomes greatly enhanced in terms of $\alpha(|\psi_i|^2)$, the amplitudes used for the histogram are those obtained from a coarse-graining procedure of the state using boxes of linear size $l=5$. Therefore the system size in all equations is the {\em effective} system size $L/l$.
The uncertainty of the PDF value is estimated from the usual standard deviation for a counting process as $\sigma_{\mathcal{P}_L(\alpha)}=\sqrt{\mathcal{P}_L(\alpha)/N_{w}L^d\Delta\alpha}$, where $N_w$ is the total number of states in the average, and for the $\alpha$ values a constant $\sigma_\alpha=\Delta\alpha/3$ is assigned. We note that this procedure assumes uncorrelated $\alpha$'s and hence $|\psi|^2$'s; this is only true between different disorder realizations, but not necessarily within each state.
Hence the errors of the PDF are probably somewhat underestimated and the small uncertainty of the $f(\alpha)$-values obtained from PDF scaling in Fig.\ \ref{fig-giprsys} must be interpreted carefully. There also exists another source of error difficult to quantify, namely, how much the histogram for finite $\Delta\alpha$ deviates from the real PDF when $\Delta \alpha\rightarrow 0$. In spite of this, the PDF method is easy to implement numerically and hence a valid alternative to the more demanding gIPR scaling techniques. 
%%%%%%%%%%%%%%%%%%%%%%%%%%%%%%%%%%%%%%%%%%%%%%%%%%%%%%%%%%%%%%%%%%%%%%%%%%
%%%%%%%%%%%%%%%%%%%%%%%%%%%%%%%%%%%%%%%%%%%%%%%%%%%%%%%%%%%%%%%%%%%%%%%%%%

\begin{figure}
  \centering
   \includegraphics[width=\figwidth]{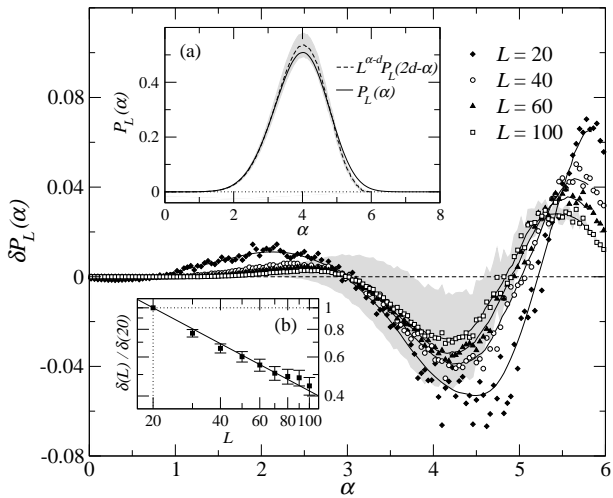}
   \caption{Degree of symmetry of the PDF using $\Delta\alpha=0.04$ for different system sizes with $2.5\times 10^4$ critical wavefunctions for each size. The shaded region in the background marks the $66\%$ c.i. of the values for $L=100$. The lines show the evolution in average for each $L$. 
   The inset (a) displays the symmetry transformed PDF for $L=100$ and its 95\% c.i. (shaded region in the background).
   The inset (b) shows a log-log plot of the cumulative difference $\delta(L)$ normalized to its maximum value (minimum L). The solid line corresponds to a linear fit with slope $b= -0.537\pm0.016$.}
\label{fig-degsymPDF}
\end{figure}

%%%%%%%%%%%%%%%%%%%%%%%%%%%%%%%%%%%%%%%%%%%%%%%%%%%%%%%%%%%%%%%%%%%%%%%%%%
\paragraph{Symmetry relation for the PDF.}
%%%%%%%%%%%%%%%%%%%%%%%%%%%%%%%%%%%%%%%%%%%%%%%%%%%%%%%%%%%%%%%%%%%%%%%%%%
The symmetry relation, $f(2d-\alpha)=f(\alpha)+d-\alpha$ \cite{MirFME06} for $L\rightarrow \infty$, implies the existence of a symmetry also for the PDF which should hold for large enough system sizes,
\begin{equation}
  \mathcal{P}_L(2d-\alpha)= L^{d-\alpha} \mathcal{P}_L(\alpha).
  \label{eq:sympdfalpha}
\end{equation}
In terms of the wavefunction amplitudes, Eq.~\eqref{eq:sympdfalpha} reads $\mathcal{P}_L(L^{-2d}/|\psi|^{2})=(L^d |\psi|^2)^3 \mathcal{P}_L(|\psi|^2)$. 
The latter relation establishes that at criticality the distribution in the interval $L^{-2d}<|\psi|^2\leqslant L^{-d}$ is indeed determined by the PDF in the region $L^{-d}\leqslant |\psi|^2 <1$.  We carry out a numerical check of the symmetry relation \eqref{eq:sympdfalpha} by evaluating $\delta\mathcal{P}_L(\alpha)=\mathcal{P}_L(\alpha)-L^{\alpha-d} \mathcal{P}_L(2d-\alpha)$ accounting for the distance between the original PDF and its symmetry-transformed counterpart at every $\alpha$, as well as the cumulative difference $\delta(L)=\int_0^{2d} d\alpha |\delta\mathcal{P}_L(\alpha)|$. The symmetry-transformed PDF for $L=100$ and the evolution of $\delta\mathcal{P}_L(\alpha)$ for different $L$ are shown in Fig.~\ref{fig-degsymPDF}. 
We find that the symmetry relation is better satisfied as $L$ increases \cite{VasRR08a, RodVR08}, and the improvement can be roughly quantified as $\delta(L)\sim L^{-0.537}$ as shown in Fig.~\ref{fig-degsymPDF}(b). 
%%%%%%%%%%%%%%%%%%%%%%%%%%%%%%%%%%%%%%%%%%%%%%%%%%%%%%%%%%%%%%%%%%%%%%%%%%

\begin{figure}
  \centering
   \includegraphics[width=.9\columnwidth]{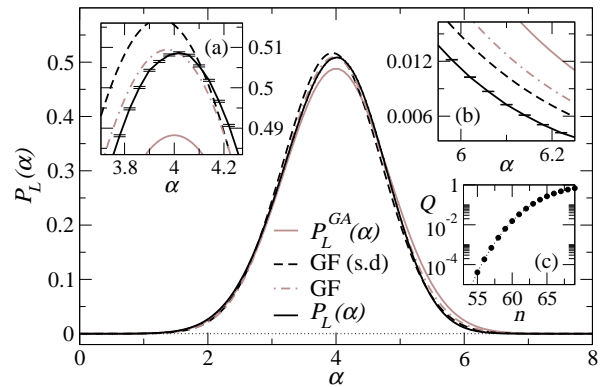}
   \caption{(color online) Gaussian fit ($ \sqrt{a/\pi}\,e^{-a(x - b)^2}$) of the PDF obtained from $2.5\times 10^4$ critical states of size $L^3=100^3$ using $\Delta\alpha=0.04$. The dashed-dot line corresponds to the fit without taking into account the uncertainty of the PDF values ($a=0.815\pm0.004,\,b=3.9731\pm0.0021 $). The dashed line shows the fit obtained when the uncertainty of the PDF points is considered ($a=0.840\pm0.005,\,b=3.942\pm0.004,\,Q<10^{-160000}$). The solid gray line shows the GA for $\epsilon=1$. Insets (a) and (b) are blow-ups of the maximum and the right tail of the PDF. Inset (c) shows the quality-of-fit parameter $Q$ when the standard deviation of the PDF-points is multiplied by the factor $n$.}
\label{fig-Nongauss}
\end{figure}

%%%%%%%%%%%%%%%%%%%%%%%%%%%%%%%%%%%%%%%%%%%%%%%%%%%%%%%%%%%%%%%%%%%%%%%%%%
\paragraph{Non-gaussian nature of the PDF.}
%%%%%%%%%%%%%%%%%%%%%%%%%%%%%%%%%%%%%%%%%%%%%%%%%%%%%%%%%%%%%%%%%%%%%%%%%%
The parabolic approximation for $f(\alpha)$ in $d=2+\epsilon$, \cite{Weg89} implies a gaussian approximation (GA) for the PDF, $\mathcal{P}^{GA}_L (\alpha) = \sqrt{\ln L/4\pi\epsilon}\, L^{-[\alpha-(d+\epsilon)]^2/4\epsilon}$. At first glance, the PDFs in Fig.~\ref{fig-multipdf}, 
 might indeed appear roughly gaussian and in Fig.~\ref{fig-Nongauss} we show gaussian fits of the PDF for $L=100$, obtained via a usual $\chi^2$ minimization taking into account the uncertainties of the PDF values. However, the quality-of-fit parameter $Q$, which gives an indication on the reliability of the fit, is ridiculously small ($Q<10^{-160000}$). Since the individual standard deviations of the PDF values may have been slightly underestimated we also study how $Q$ behaves when we intentionally increase the error bars by a factor $n$. As shown in Fig.~\ref{fig-Nongauss}(c) one would have to go to unreasonable high values of $n \sim 60$ to accept a gaussian nature for the PDF as plausible. The deviation from $\mathcal{P}_L^{GA}(\alpha)$ is also noticeable. Hence our statistical analysis confirms that the PDF is non-gaussian in agreement with the observed non-parabolic nature of $f(\alpha)$ at the 3D MIT \cite{VasRR08a,RodVR08}. %%%%%%%%%%%%%%%%%%%%%%%%%%%%%%%%%%%%%%%%%%%%%%%%%%%%%%%%%%%%%%%%%%%%%%%%%%
%%%%%%%%%%%%%%%%%%%%%%%%%%%%%%%%%%%%%%%%%%%%%%%%%%%%%%%%%%%%%%%%%%%%%%%%%%

\begin{figure}
  \centering
   \includegraphics[width=.49\columnwidth]{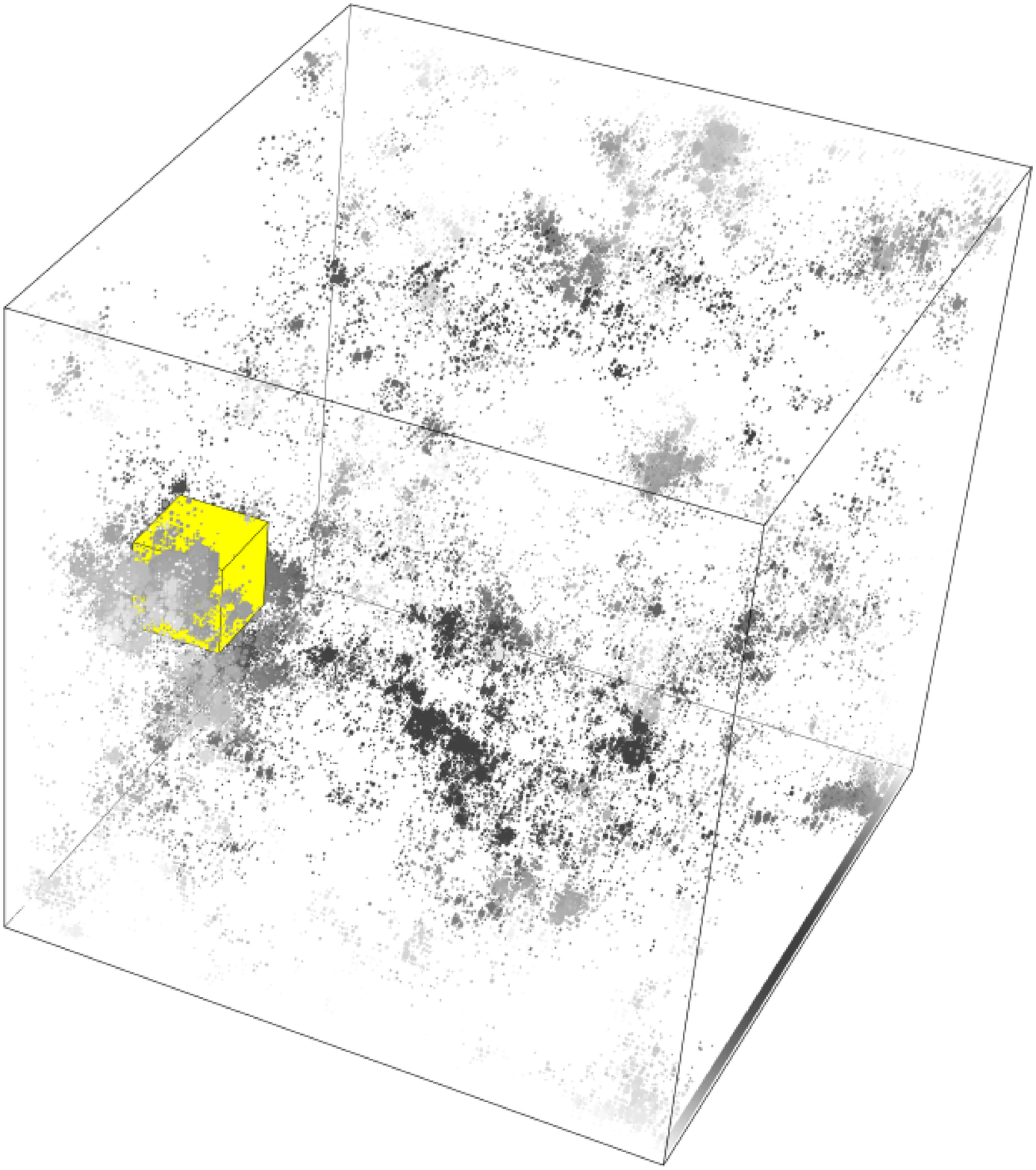}
   \includegraphics[width=.49\columnwidth]{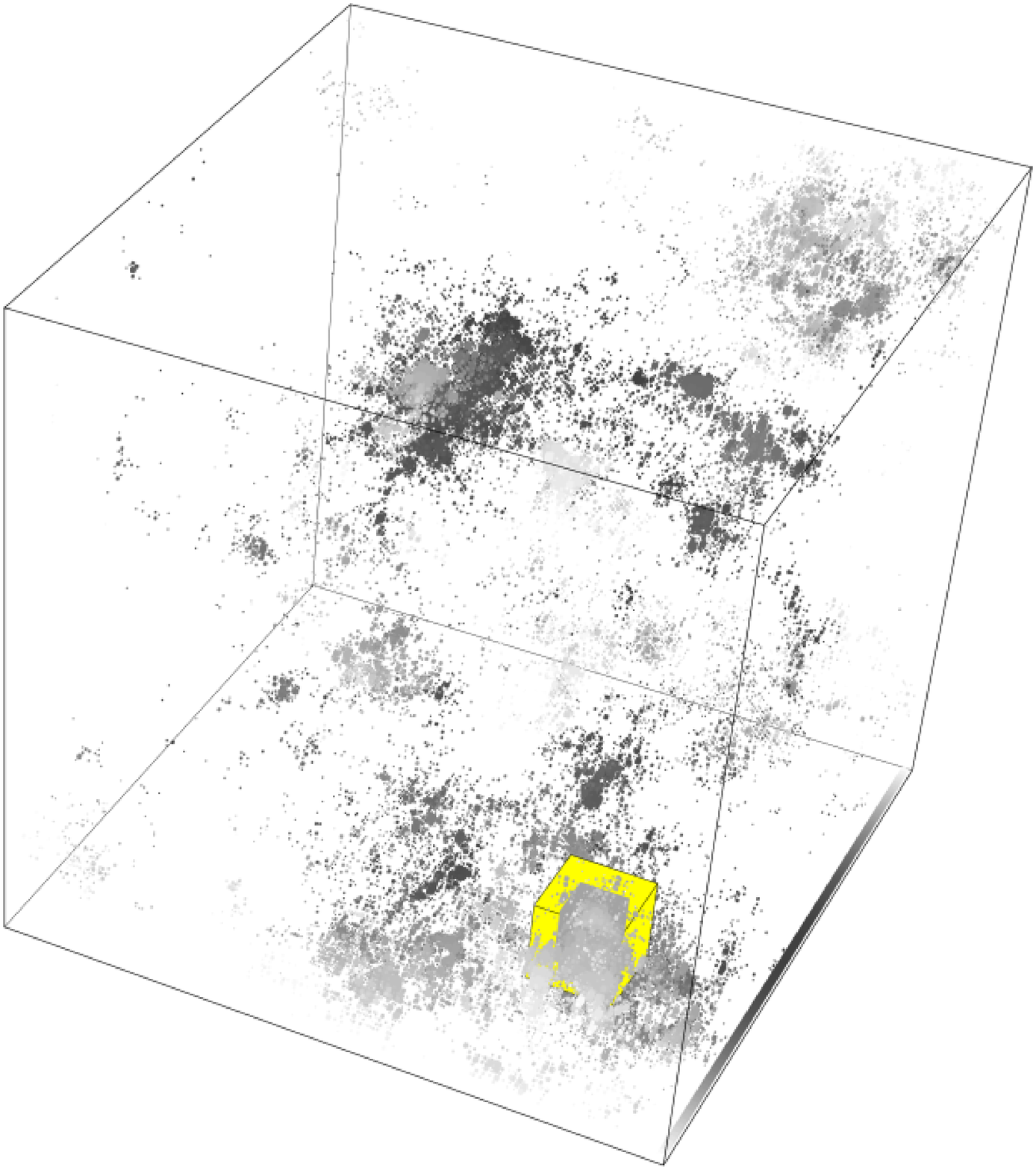}
   \caption{(color online) Rare eigenstates for the 3D Anderson model at $E=0$ and $W_c=16.5$ for $L=100$.  The sites with probability $|\psi_j|^2$ larger than $L^{-3}$ are shown as boxes with volume $|\psi_j^2| L^3$. The grayscale distinguishes between different slices of the system along the axis into the page. The biggest boxes with black edges enclose the site with the maximum normalized amplitude: $\vert\psi_i\vert^2=0.4484$ (left) and $0.3617$ (right).}
\label{fig-rarestate}
\end{figure}

\begin{figure}
  \centering
   \includegraphics[width=\figwidth]{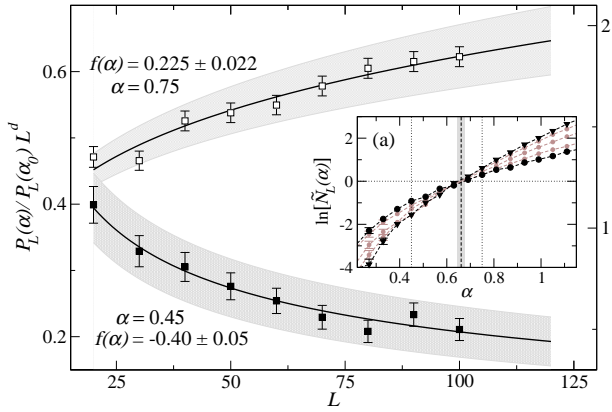}
   \caption{(color online) $\widetilde{N}_L(\alpha)$ vs $L$ for $\alpha=0.45$ (filled squares, left $y$ axis) and $0.75$ (open squares, right $y$ axis).    
   The solid lines are the $L^{f(\alpha)}$ curves where $f(\alpha)$ is obtained from the log-log linear fits of Eq.~\eqref{eq:correctedscaling}. The shaded regions mark the $66\%$ c.i. of the $L^{f(\alpha)}$ values.  The inset shows $\ln[\widetilde{N}_L(\alpha)]$ vs $\alpha$ for different $L$ in the range $[20\; (\textrm{black circle}),100\;(\textrm{black triangle})]$.  Standard deviations are within symbol size, whenever not shown. The vertical dashed line corresponds to the mean value of the crossing point $\alpha_-= 0.661\pm0.008$ and the shaded bar marks its $95\%$ c.i..}
\label{fig-NvsL}
\end{figure}

%%%%%%%%%%%%%%%%%%%%%%%%%%%%%%%%%%%%%%%%%%%%%%%%%%%%%%%%%%%%%%%%%%%%%%%%%%
\paragraph{Rare events and their negative fractal dimensions.}
%%%%%%%%%%%%%%%%%%%%%%%%%%%%%%%%%%%%%%%%%%%%%%%%%%%%%%%%%%%%%%%%%%%%%%%%%%
The volume of the $\alpha$-set, $\mathcal{N}_L(\alpha)$ for a given $L$, gives the number of points in the wavefunction with amplitudes in the range $\vert\psi_i\vert^2\in[L^{-\alpha-\Delta\alpha/2},L^{-\alpha+\Delta\alpha/2}]$. It scales with the system size as $\mathcal{N}_L(\alpha) \sim \sqrt{\ln L}\, L^{f(\alpha)}$. The negative values of $f(\alpha)$ \cite{RodVR08} correspond then to those $\alpha$-sets whose volume decreases with $L$ for large enough $L$. 
Physically, the negative fractal dimensions at small $\alpha$ are caused by the so-called rare events containing localized-like regions of anomalously high $\vert\psi_i\vert^2$ at criticality. The probability of finding them likewise decreases with $L$.  In Fig.~\ref{fig-rarestate}, we show examples of rare eigenstates. 
%%%%%%%%%%%%%%%%%%%%%%%%%%%%%%%%%%%%%%%%%%%%%%%%%%%%%%%%%%%%%%%%%%%%%%%%%%
%%%%%%%%%%%%%%%%%%%%%%%%%%%%%%%%%%%%%%%%%%%%%%%%%%%%%%%%%%%%%%%%%%%%%%%%%%
Due to the finite size term $\sqrt{\ln L}$, the threshold $\alpha_-$ [where $f(\alpha_-)=0$], below which the decreasing behaviour of $\mathcal{N}_L(\alpha)$ with $L$ is detected, will change with the system size itself, and so the normalized volume \eqref{eq:correctedscaling} of the $\alpha$-set must be used. In Fig.~\ref{fig-NvsL} we show the behaviour of $\widetilde{\mathcal{N}}_L(\alpha)$ vs $L$ for two values of $\alpha$ corresponding to a positive and a negative fractal dimension. The exponential decreasing of the volume of the $\alpha$-set for $f(\alpha)<0$ is clearly observed from the PDF values. In Fig.~\ref{fig-NvsL}(a) it is demonstrated how the normalized volume of the $\alpha$-set becomes scale invariant at $\alpha_-$ and thus  $\widetilde{\mathcal{N}}_L(\alpha)=1$. The opposite tendencies with $L$ at each side of $\alpha_-$ can also be seen.
The estimated value for $\alpha_-\in[0.644,0.677]$ from the PDF scaling agrees with the result obtained using the multifractal spectrum from gIPR scaling, 
$\alpha_-=0.626\pm0.028$ \cite{RodVR08}.
%%%%%%%%%%%%%%%%%%%%%%%%%%%%%%%%%%%%%%%%%%%%%%%%%%%%%%%%%%%%%%%%%%%%%%%%%%
%%%%%%%%%%%%%%%%%%%%%%%%%%%%%%%%%%%%%%%%%%%%%%%%%%%%%%%%%%%%%%%%%%%%%%%%%%

\begin{figure}
  \centering
   \includegraphics[width=.9\columnwidth]{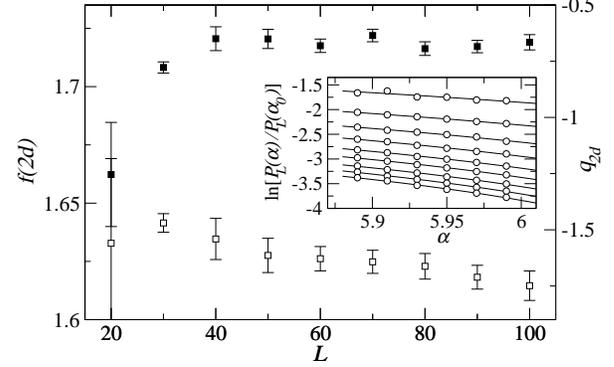}
   \caption{Values of $f(2d)$ (filled squares, left $y$ axis) and $q_{2d}\equiv f'(\alpha)|_{\alpha=2d}$ (open squares, right $y$ axis) as functions of system size. The inset shows the fits of Eq.~\eqref{eq-pdfseries2d} for different $L$ from $20$ (top) to $100$ (bottom). Standard deviations are within symbol size.}
\label{fig-termination}
\end{figure}

%%%%%%%%%%%%%%%%%%%%%%%%%%%%%%%%%%%%%%%%%%%%%%%%%%%%%%%%%%%%%%%%%%%%%%%%%%
\paragraph{Termination points in $f(\alpha)$.}
%%%%%%%%%%%%%%%%%%%%%%%%%%%%%%%%%%%%%%%%%%%%%%%%%%%%%%%%%%%%%%%%%%%%%%%%%%
The fate of the $f(\alpha)$ spectrum at $\alpha=0$ and $\alpha=2d$ is currently under debate in the literature \cite{ObuSFGL07,RodVR08} due to the emergence of singularities at these points. At present, it is not clear whether $f(\alpha)$ continues towards $-\infty$ or terminates with finite values. 
The physical consequences of the absence of termination points (TP) are: (i) $|\psi_i|^2>L^{-2d}$ at criticality since $2d$ would be an upper bound for $\alpha$, and (ii) the probability to find the most rare event, namely the most extremely localized state ($|\psi_{i_0}|^2=1$, corresponding to $\alpha=0$), at the critical point must always be zero independently of the system size [$\mathcal{P}_L(0)=0$]. 
In principle the PDF can be used to look for TPs both at $\alpha=0$ and $2d$. However, a reliable analysis in the vicinity of $\alpha=0$ requires a huge number of disorder realizations; relying on the symmetry relation \cite{MirFME06} a study around $\alpha=2d$ is more appropriate.
For $1\ll\alpha<2d$ and as long as there is a TP, the PDF admits the series expansion 
 \begin{equation}
  \mathcal{P}_L(\alpha)\simeq\mathcal{P}_L(\alpha_0)L^{f(2d)-d} 
\left[1+q_{2d} (\alpha-2d) \ln L\right],
\label{eq-pdfseries2d}
 \end{equation} 
where $q_{2d}\equiv f'(\alpha)|_{\alpha=2d}$. The existence of a TP requires $f(2d)$ and $q_{2d}$ to be $L$-independent, for large enough $L$. 
In Fig.~\ref{fig-termination} we show the values of $f(2d)$ and $q_{2d}$ obtained for different $L$. The value of $f(2d)$  seems to reach a saturation for large $L$ although the numerical analysis cannot exclude a very slow decreasing tendency. A similar result is found for $q_{2d}$. Still larger $L$ and more states are needed to decide the fate of $f(\alpha)$ at $0$ and $2d$. 
%%%%%%%%%%%%%%%%%%%%%%%%%%%%%%%%%%%%%%%%%%%%%%%%%%%%%%%%%%%%%%%%%%%%%%%%%%
%%%%%%%%%%%%%%%%%%%%%%%%%%%%%%%%%%%%%%%%%%%%%%%%%%%%%%%%%%%%%%%%%%%%%%%%%%

In conclusion,  
we have shown here that a PDF-based study can be a valid companion approach to the standard multifractal analysis, giving complementary and new information about the critical properties of waves at Anderson-type transitions as well as offering a conceptually simpler viewpoint.

%%%%%%
\begin{acknowledgements}
 A.R. acknowledges financial support from the Spanish government (FIS2006-00716, MMA-A106/2007) and JCyL (SA052A07). R.A.R.\ gratefully acknowledges EPSRC (EP/C007042/1) for financial support.
\end{acknowledgements}
%%%%%%%%%%%%%%%%%%%%%%%%%%%%%%%%%%%%%%%%%%%%%%%%%%%%%%%%%%%%%%%%%%%%%%%%
%%%%%%%%%%%%%%%%%%%%%%%%%%%%%%%%%%%%%%%%%%%%%%%%%%%%%%%%%%%%%%%%%%%%%%%%
%  References
%%%%%%%%%%%%%%%%%%%%%%%%%%%%%%%%%%%%%%%%%%%%%%%%%%%%%%%%%%%%%%%%%%%%%%%%
%\bibliographystyle{prsty}\bibliography{bibliography/bibliograph}

\end{document}